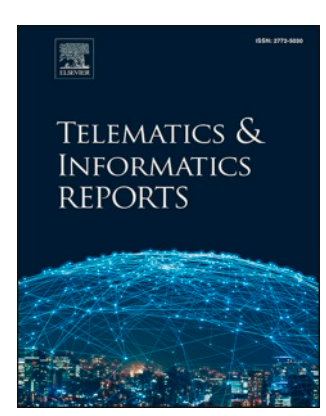

# Understanding human-AI trust in education

Griffin Pitts [a,*], Sanaz Motamedi [b]

[a] Department of Computer Science, NC State University, Raleigh, NC, USA
[b] Department of Industrial and Manufacturing Engineering, The Pennsylvania State University, University Park, PA 16802, USA

ABSTRACT

As AI chatbots become integrated in education, students are turning to these systems for guidance, feedback, and information. However, the anthropomorphic characteristics of these chatbots create ambiguity over whether students develop trust in them in ways similar to trusting a human peer or instructor (human-like trust, often linked to interpersonal trust models) or in ways similar to trusting a conventional technology (system-like trust, often linked to technology trust models). This ambiguity presents theoretical challenges, as interpersonal trust models may inappropriately ascribe human intentionality and morality to AI, while technology trust models were developed for non-social systems, leaving their applicability to conversational, human-like agents unclear.

To address this gap, we examine how these two forms of trust, human-like and system-like, comparatively influence students' perceptions of an AI chatbot, specifically perceived enjoyment, trusting intention, behavioral intention to use, and perceived usefulness. Using partial least squares structural equation modeling, we found that both forms of trust significantly influenced student perceptions, though with varied effects. Human-like trust was the stronger predictor of trusting intention, whereas system-like trust more strongly influenced behavioral intention and perceived usefulness; both had similar effects on perceived enjoyment.

The results suggest that interactions with AI chatbots give rise to a distinct form of trust, *human-AI trust*, that differs from human-human and human-technology models, highlighting the need for new theoretical frameworks in this domain. In addition, the study offers practical insights for fostering appropriately calibrated trust, which is critical for the effective adoption and pedagogical impact of AI in education.

## 1. Introduction

Trust, a psychological construct with varying definitions across disciplines and contexts, has become an important factor in understanding student interactions with artificial intelligence (AI) in education. Students may rely on AI systems for guidance, feedback, and information under conditions of uncertainty. In education, trust has been found to influence students' engagement, perceptions, and outcomes across student-student [1], student-instructor [2,3], student-institution [4,5], and student-technology relationships [6–9]. As educational technologies have become central to students' learning experiences, understanding the development of trust between students and these systems has become important for their effective use and adoption [6–9]. When trust formation is not well understood in the design and development of AI systems, students may develop inappropriate levels of trust, either over-trusting AI systems and becoming overly reliant on potentially inaccurate information, or under-trusting these systems and rejecting helpful educational tools altogether [10]. Such misaligned trust can lead to poor learning outcomes, reduced engagement with beneficial technologies [9], and the ineffective integration of AI in education [7,8,11].

In recent years, AI has become more available and used in educational settings, such as with AI chatbots. These programs, often powered by large language models (LLMs), can engage in human-like text conversations and have been used across cases for teaching and learning [12–15], administration, student assessment, and advisory purposes in educational settings [14]. For instance, AI chatbots have been developed as virtual teaching assistants that provide personalized tutoring and discussion [16]. In computing education, they have been explored as pair programmers capable of generating explanations and examples [17,18] and as tools for feedback on code quality, such as identifying misleading variable names and suggesting improvements [19], potentially substituting for peers or instructors during the learning process. Student adoption of these technologies has been substantial, with recent research finding that 64.5 % of surveyed undergraduate students use AI chatbots at least once a week, with 90.4 % reporting prior experience with these systems [20].

* Corresponding author.
*E-mail addresses:* wgpitts@ncsu.edu (G. Pitts), sjm7946@psu.edu (S. Motamedi).

Students' interactions with AI chatbots often resemble exchanges with peers or instructors, as students receive personalized assistance through simulated human dialogue. This resemblance creates challenges for how trust should be measured and modeled. Features such as natural language conversation, responsiveness, and the ability to mimic social cues blur whether students' trust aligns more with interpersonal frameworks, typically applied to human relationships [21], or with technology trust frameworks designed for human-technology interactions [22]. These characteristics may lead students to treat chatbots as if they were social partners, attributing qualities like benevolence or integrity, while also evaluating them as tools in terms of reliability, consistency, and usefulness.

Traditional models offer limited guidance in this area. Interpersonal trust frameworks designed for human-human interactions assume attributes such as benevolence and integrity, which require reinterpretation when applied to AI systems that lack genuine intention or agency [22]. Conversely, technology trust frameworks designed for human-technology relationships have been developed and validated primarily for non-social systems, raising questions about their applicability to technologies with more anthropomorphic, or human-like, features (e.g. simulated conversation) [23,24].

Building on this recognition, Lankton et al [24] proposed distinguishing between human-like and system-like trusting beliefs in technologies. Human-like trust reflects perceptions of ability, benevolence, and integrity, attributes typically associated with interpersonal relationships. System-like trust, by contrast, is based around perceptions of functionality, helpfulness, and reliability. Prior research has shown that the relative influence of these beliefs depends on context. Trust in applications like Facebook and Airbnb has been validated to have a stronger influence from human-like trusting beliefs than system-like trusting beliefs, hypothesized to be based around their affordances for social connection and interpersonal communication [24,25]. Whereas trust in tools like Microsoft Excel has been validated to have a greater influence from system-like trusting beliefs than human-like trusting beliefs, hypothesized to be related to their lack of social or human-like affordances [24]. This distinction parallels other theoretical perspectives, such as cognitive versus affective trust [26] and performance-based versus moral trust [27]. Yet uncertainty remains over which framework is most appropriate for measuring trust in AI systems, creating challenges for researchers and technology developers seeking to understand and predict user trust, risking inappropriate trust calibration, suboptimal system design, and failed technology adoption.

To address this gap, this study investigates the relative importance of human-like and system-like trust in students' interactions with AI chatbots for learning support. We specifically examine how these two forms of trust influence four outcome variables: perceived usefulness, enjoyment, trusting intention, and behavioral intention to use an AI assistant. These outcomes were selected following Lankton et al [24], who demonstrated that trust in technology shapes general behavioral beliefs (usefulness and enjoyment), object-oriented attitudes (trusting intention), and intentions to continue using a technology (behavioral intention). Given the social affordances of AI chatbots (e.g. their capability to simulate human dialogue), we hypothesize that human-like trust will exert stronger influence on the outcome variables than system-like trust, despite students' awareness that they are interacting with a software-based system. To test this hypothesis, we applied partial least squares structural equation modeling (PLS-SEM) to surveyed data from 229 undergraduate students. The findings extend understanding of how trust develops in student-AI interactions and support the view that human-AI trust constitutes a distinct form, differing from both human-human and human-technology relationships.

In the remainder of the paper, a review of relevant literature relating to trust is provided, focusing on the conceptualization and measurement of prior models of interpersonal and technological trust. We then consider how anthropomorphism, social affordances, and the CASA paradigm complicate the application of these models to AI chatbots, leading to the theoretical gap in understanding student-AI trust. Building on this background, we develop our hypotheses, outline the research methodology, present and analyze the findings, and conclude with implications, limitations, and directions for future work.

## 2. Literature review

This literature review examines the theoretical foundations underlying trust. First, we discuss established definitions of trust and distinguish it from related constructs, then review interpersonal trust (e.g., [21]) and technology trust models (e.g., [22]). We consider trust in education across student-student, student-instructor, and student-institution relationships as a lens for student-AI trust. Finally, we examine how anthropomorphism affects measurement and conceptualization, motivating the gap our study addresses.

### *2.1. Conceptual background on trust*

Across disciplines, trust lacks a single acceptance definition, remaining largely dependent on the context and discipline in which it is applied [22,28,29]. Generally, trust is a subjective and psychological construct that enables cooperation, reduces uncertainty, and facilitates interactions between individuals, groups, and systems [21,22,30]. Common formulations emphasize *vulnerability/risk* [21,31] or *confidence in reliability/ability* [32], and in human-automation contexts an *attitude that an agent will help achieve goals under uncertainty* [33]. Although given in different contexts, these definitions share a common theme in which trust involves an expectation that the trustee (the person, group, or technology being trusted) will act consistently with the trustor's interests.

Beyond trust itself, we distinguish trust (a trustor's belief/willingness to rely) from trustworthiness (properties of the trustee) [34], and note that distrust is not merely low trust but a distinct negative expectation that can coexist with trust [10,35–37]. In educational settings, imbalances in either direction, over-trust or over-distrust, can influence how AI in education is perceived and engaged with, leading to varying degrees of acceptance toward these tools from students and instructors [6,8,38–46]. A balanced approach involving appropriate levels of trust and distrust is optimal, while extremes in either direction can be problematic [10].

For the purposes of this study, we define student-AI trust as "*a student's willingness to rely on an AI system's guidance, feedback, and information under conditions of uncertainty and potential vulnerability*". This definition acknowledges the varying roles of AI systems in education, and the uncertainties and potential vulnerability students' face in AI interactions, relating to system accuracy and reliability, and risks to privacy, learning outcomes, and intellectual growth [47–52]. Depending on the pedagogical role of an AI chatbot, student-AI trust can be an influential factor as students decide whether to use AI-generated content, provide personal information to AI systems, or follow AI-generated learning recommendations. As these roles parallel traditional student-student and student-instructor, it is necessary to first consider how interpersonal trust has been modeled before examining its applicability to human-AI relationships.

### *2.2. Interpersonal trust*

Interpersonal trust refers to a form of trust that develops between people, whether as individuals, groups, or organizations [53]. Several theoretical frameworks have been developed to understand interpersonal trust, each offering different perspectives on how trust forms and functions. Mayer et al.'s [21] model of organizational trust, one of the most widely cited in trust research, identified three factors that influence interpersonal trustworthiness: perceived ability, benevolence, and integrity. Other interpersonal trust models have explored different perspectives, such as McAllister's [26] distinction between cognition-and affect-based trust or Lewicki and Bunker's [54] sequential stages of trust development. Although these models emphasize different mechanisms, interpersonal trust plays a particularly important role in educational settings, where students' trust in peers, instructors, and institutions directly influences collaboration, engagement, and learning outcomes. Given the growing role of AI systems as collaborators, tutors, or institutional agents, understanding how interpersonal trust functions across these educational relationships can provide insight into the potential impacts of student trust on engagement, motivation, and performance when similar roles are taken on by AI.

#### *2.2.1. Student-student trust, student-instructor trust, student-institution trust*

Student-student trust refers to the trust established between peers in educational settings. It has been found to influence engagement in collaborative learning environments, including in group projects [1], peer assessment [55], and peer learning [56,57]. Such trust influences students' willingness to share ideas, participate in discussions, and reflect critically on their own and others' viewpoints [58–60].

Student-instructor trust refers to the trust developed between students and their teachers or professors, developed through an initial impression and a continuing pattern of interactions [3]. Student-instructor trust can play a role in classroom interactions, office hours discussions, and in assignment feedback. The quality of and trust involved in student-instructor interactions has influence toward academic performance [3,61,62], with higher levels of trust between students and instructors correlating with increased classroom engagement [59,61,63], academic motivation [59,64], and course satisfaction [2].

Student-institution trust refers to the trust students place in their educational institutions, with implications toward student retention and satisfaction with educational experiences [65–67]. When students trust their institutions, they are more likely to engage with available resources, adhere to institutional policies, and develop a stronger affiliation with the broader educational community [65–67].

As AI systems take on roles resembling peers, instructors, or institutions, students may extend similar trust dynamics to these technologies, with consequences for engagement, motivation, and performance. However, the transferability of established interpersonal trust frameworks to AI systems remains an open question, particularly as these systems continue to advance in sophistication and autonomy. This uncertainty raises the question of whether students apply interpersonal trust models when interacting with AI, or whether technology trust models provide a better foundation for understanding these relationships.

### *2.3. Technology trust*

While trust has traditionally been studied in human-human or human-organizational relationships, as referred to in the previous section, the development of digital technologies necessitated a reconceptualization of trust theory. Researchers have argued that applying interpersonal trust frameworks to technological contexts presents conceptual limitations [22]. Interpersonal models, such as Mayer et al.'s [21] framework of ability, benevolence, and integrity, assume that the trustee possesses moral agency, intentionality, and reciprocity. However, as McKnight et al [22] point out, technologies "cannot 'care' about the trustor or have 'integrity' because they lack volition". This critique raised questions about whether people genuinely experience trust in technologies in the same way they trust humans, or whether different psychological processes are involved.

Recognizing these conceptual challenges, research has established that people do form trusting relationships with technologies, though these relationships exhibit distinctive characteristics compared to interpersonal trust [22]. To account for these differences, McKnight et al [22] developed a technology-specific model that identifies functionality, helpfulness, and reliability as the primary determinants of trust in technological systems. Further, studies have explored trust in specific technologies, such as Gulati et al [23] who found willingness, ability, benevolence, and reciprocity to be significant factors in individuals' trust in Siri, suggesting that trust in a technology can draw on both performance-oriented and socially-oriented judgments. The advent of artificial intelligence introduced additional complexity to this domain. Specific to AI technologies, researchers have proposed new frameworks that account for the unique characteristics of these systems. Siau and Wang [68] proposed that trust in AI incorporates not only technological characteristics of performance, process, and purpose, but also environmental characteristics such as task nature, cultural background, and institutional factors, as well as human characteristics relating to individuals' personality and disposition to trust. Jacovi et al [69] formalized human-AI trust as *contractual trust*, where reliance on an AI system rests on the expectation that an implicit or explicit contract will hold, rather than on assumptions of human-like intentionality. Trust modeling specific to pedagogical conversational AI has been explored, with Pitts et al [70] proposing a model of learners' acceptance and trust that includes factors relating to individuals' propensity to trust and perceived AI competence, benevolence, privacy risk, and information quality risk. Scholars and policymakers have also proposed normative frameworks for 'trustworthy AI,' emphasizing technical qualities like robustness and transparency alongside value-based considerations such as fairness, benevolence, and integrity [71,72]. These overlaps suggest that human-AI trust may draw simultaneously on performance-oriented and value-oriented judgments, integrating elements of both technology and interpersonal trust.

While much of this work has examined human-technology trust in general contexts, its implications are highly relevant in education, as students must regularly decide how much to rely on, engage with, and follow the recommendations of AI systems. When trust is miscalibrated, the consequences can include poorer learning outcomes, reduced long-term engagement, and reluctance to adoption of potentially beneficial AI-enhanced educational technologies [8,11].

#### *2.3.1. Student-technology trust*

In educational settings, prior work has shown that higher levels of student-technology trust can be associated with increased adoption, more frequent use, and greater willingness to follow system recommendations [6–9]. For example, with automated feedback systems, trust influences the extent to which students carefully consider and apply feedback to improve their work [9]. With e-authentication systems for online assessments, trust can influence whether students accept educational technologies as secure or perceive it as increased institutional distrust of their academic integrity [73]. In online learning environments, trust can promote enrollment, reduce dropout rates, strengthen student-instructor relationships [74]. The importance of trust becomes evident when considering that positive engagement with educational technologies is closely tied to improved learning outcomes, whereas low trust often leads to disengagement, frustration, and avoidance [9,75]. Yet it remains unclear how the social and anthropomorphic features of AI systems influence the ways students develop and sustain trust, raising questions about whether existing technology trust frameworks are sufficient for these emerging educational tools. This makes it important to examine anthropomorphism, as many AI systems incorporate human-like features, such as natural language conversation, that create affordances for social connection that may influence how students perceive and place trust in them.

### *2.4. Anthropomorphism and AI in education*

Anthropomorphism refers to the attribution of human characteristics, behaviors, or emotions to non-human entities, including technologies [76,77]. In technological contexts, anthropomorphism manifests when users perceive or interact with technologies as if they possess human-like qualities, such as intentions, feelings, or consciousness [77].

This tendency to anthropomorphize occurs naturally, serving as a cognitive mechanism that helps individuals understand, predict, and relate to non-human agents by applying familiar social schemas [78]. The degree of anthropomorphism varies based on several factors, including the technology's design features, behavioral cues, and contextual factors [79]. Design characteristics that promote anthropomorphism include human-like appearance, natural language processing, personalized responses, conversational turn-taking, and expressions of apparent emotion or personality [80,78,81]. Additionally, individual differences, such as prior technological experiences and cultural backgrounds, can influence individual's propensity to anthropomorphize technologies [82].

The development and use of anthropomorphic technologies in educational settings has increased in recent years, with interfaces ranging from AI chatbots [14,29,83], to educational robots [84–87], and virtual agents [88,89]. Among these, AI chatbots represent one of the most prominent forms, offering human-like interaction through natural language processing, personalized responses, and conversational turn-taking. Research suggests that such features can increase student engagement and motivation [90] and foster social presence effects comparable to human interactions [91]. At the same time, excessive or poorly designed anthropomorphism can have negative consequences, such as the discomfort associated with the 'uncanny valley' [92,93]. Developers must also consider the degree of anthropomorphism in designing for trust repair, since human-like agents require different strategies for acknowledging limitations [80,94].

To better understand why anthropomorphic design features influence trust formation, two theoretical perspectives can provide insight. Affordance theory describes how technologies create opportunities for both functional and social interaction, while the Computers Are Social Actors (CASA) paradigm explains why people may respond to these cues as if they were engaging with another human.

#### 2.4.1. Affordance theory and the computers are social actors (CASA) paradigm

Affordance theory defines affordances as "opportunities for action" that arise from the attributes of technologies [95]. Gibson [95] distinguished between affordances of inanimate objects and those of humans or animals, a distinction that later researchers extended to differentiate object affordances from social affordances [96]. Object affordances relate to functional attributes that enable action [95], like a door handle affording pulling or pushing based on its physical shape [97], or an AI chatbot's text input field affording typing in a message and its send button affording submission of that message. These affordances are relatively static and consistent across uses.

In contrast, social affordances reflect possibilities for social engagement and interpersonal communication [96]. Technologies can provide social affordances by appearing and behaving in human-like ways. For example, social robots present multiple social affordances through their physical embodiment, conversational abilities, responsive behaviors, and often times, facial expressions [98]. AI chatbots offer social affordances through their conversational interfaces, ability to maintain dialogue context, use of natural language, and capacity to express empathy or personality through word choice and tone. Social affordances foster the development of emotional attachments and encourage users to apply human social norms during technology interactions, facilitating interpersonal trust formation.

The CASA paradigm offers a complementary theoretical lens for explaining why these effects occur. Established by Nass et al [99], CASA demonstrates that "individuals' interactions with computers are fundamentally social" and that people apply interpersonal norms even while fully aware they are engaging with machines [99,100]. Whenever technologies exhibit human-like behaviors such as words for output, interactivity based on multiple prior inputs, and the filling of roles traditionally filled by humans, users are more likely to treat them as social actors [78]. This has been observed in behaviors such as politeness [101], gender stereotyping [102], personality attribution [103], and even responding positively to flattery [104].

Together, affordance theory and the CASA paradigm provide complementary perspectives for understanding how the anthropomorphic features of AI chatbots can create opportunities for social interaction and elicit interpersonal responses. Building on these perspectives, we propose that when chatbots display conversational abilities, empathy, or personality cues, students may be inclined to evaluate them as if they were social partners. At the same time, since chatbots are inherently software systems, students may also base their evaluations on their functionality. This raises the question of whether students' trust in chatbots aligns more closely with human-like or with system-like beliefs, a distinction developed in the next section.

### 2.5. Human-like and system-like trusting beliefs

Human-like and system-like trusting beliefs represent two perspectives for understanding how people place trust in technology. Human-like trust adapts interpersonal trust theory [21], applying constructs such as ability, *the belief that the trustee has the skills and competencies to be influential within a domain*; benevolence, *the belief that the trustee will want to do good for the trustor*; and integrity, *the belief that the trustee adheres to principles the trustor finds acceptable*. In contrast, system-like trust builds on technology-specific models [22,24], emphasizing functionality, *the belief that a technology can perform the tasks needed*; helpfulness, *the belief that the technology provides adequate and responsive support*; and reliability, *the belief that the technology will consistently operate as expected*. These perspectives parallel other dichotomies in trust research, including cognitive versus affective trust [26,105] and performance-based versus moral trust [27]. While terminology varies, each highlights how trust may develop through emotional or rational assessments depending on the context and the characteristics of the trustee.

Lankton et al [24] demonstrated that the perceived "humanness" of a technology, defined as the extent to which it is seen as more person-like than tool-like, influences whether users apply human-like or system-like trusting beliefs. Their research demonstrated that technologies with higher perceived humanness (like social networks) evoke a stronger influence from human-like trusting beliefs, while more tool-like technologies (like database software) primarily engage system-like trusting beliefs. Building on this, Califf et al [25] found that in the context of Airbnb, a platform embedded in the sharing economy with high human-to-human connection affordances, human-like trusting beliefs had significantly stronger effects on perceived usefulness, enjoyment, and continuance intention compared to system-like trusting beliefs. On the other hand, Abdulrahman Al Moosa et al [106] demonstrated that for mobile banking applications, which possess lower humanness and fewer social affordances, system-like trust was more influential than interpersonal trust for perceived usefulness and adoption intentions.

Building on prior work showing that the perceived humanness and social affordances of technologies influence whether human-like or system-like trust is more salient [24,25,106], the following section outlines the aims and hypotheses of this study.

## 3. Aims and hypothesis

This study investigates the nature of student-AI trust by examining the relative influence of human-like and system-like trusting beliefs on students' perceptions of an AI chatbot for learning support. Specifically, we explore how these two types of trusting beliefs differentially affect four outcome variables relating to student's perceived enjoyment, trusting intention, behavioral intention to use an AI assistant, and perceived usefulness, as shown in Fig. 1. These outcomes were selected in line with prior research, particularly Lankton et al [24], who demonstrated that trust-related beliefs influence general behavioral beliefs (usefulness and enjoyment), object-oriented attitudes (trusting intention), and continuance intentions related to technology use. This aligns with research on the Technology Acceptance Model (TAM) [107], where trust has been recognized as an important extension influencing perceptions of usefulness and enjoyment as well as intentions to adopt and continue using technology [108–111]. Prior work further suggests that, although outcomes may be differentiated as emotional (e.g., enjoyment, trusting intention) versus functional (e.g., usefulness, behavioral intention), trust has been shown to act as a broad antecedent influencing all of them when measured appropriately [24,25,106].

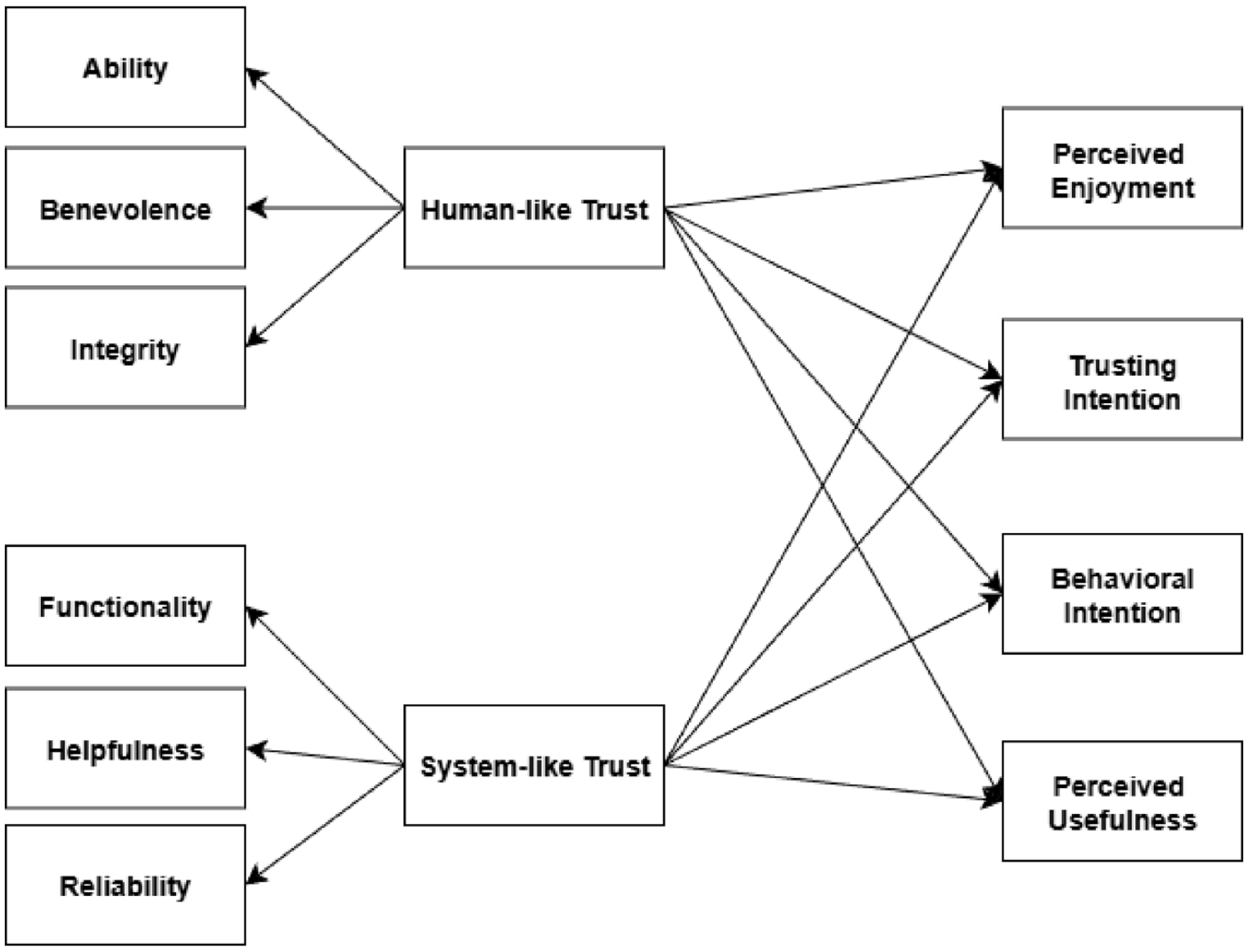

**Fig. 1.** Conceptual model of human-like and system-like trusting beliefs. Arrows indicate hypothesized causal paths from trust beliefs to outcome variables. H1-H4 predict that human-like trust will have stronger effects on each outcome than system-like trust.

It is hypothesized that human-like trust will exert stronger influence on the outcome variables than system-like trust. The rationale for these hypotheses stems from the social affordances of AI chatbots through their conversational interfaces, personalized responses, and human-like communication patterns. These affordances create opportunities for students to engage with AI systems in ways that mirror interpersonal student-student, student-instructor, or student-institution interactions. Moreover, the CASA paradigm reinforces this rationale by suggesting that students naturally respond to technologies exhibiting human-like characteristics by applying social norms and forming social judgments, even when they are consciously aware they are interacting with machines.

The following hypotheses are proposed:

H1: Human-like trust will have a stronger positive relationship with perceived enjoyment of an AI chatbot than system-like trust.

Trusting beliefs, when appropriately measured, should have a positive influence on perceived enjoyment because technologies with trustworthy attributes reduce users' feelings of risk and uncertainty, leading to greater comfort and enjoyment during use [24]. Prior research has demonstrated positive effects of trust on perceived enjoyment [24,25,112]. The anthropomorphic characteristics and conversational interactions of AI chatbots suggest human-like trust will be more influential in explaining variance in perceived enjoyment.

H2: Human-like trust will have a stronger positive relationship with trusting intention toward an AI chatbot than system-like trust.

Trusting beliefs should positively influence trusting intention because individuals with strong trusting beliefs perceive that the trustee possesses desirable characteristics that enable future trust [24,113]. While both trust types can influence willingness to depend on a technology ([25,114,115], 2015), the relative importance is expected to depend on technological humanness, with human-like trust more influential in explaining variance in trusting intention.

H3: Human-like trust will have a stronger positive relationship with behavioral intention to use an AI chatbot than system-like trust.

Recent extensions of the Technology Acceptance Model (TAM) emphasize that trust is central in reducing uncertainty and perceived risk, both of which are barriers to adoption of emerging technologies [108–111]. When users perceive a system as trustworthy, they are more willing to engage with it despite potential risks, because trust provides a psychological assurance that the technology will perform as expected [22,24]. In this sense, trust serves as a psychological mechanism that helps users reduce concerns about undesirable technology outcomes, which in turn influences behavioral intention.

H4: Human-like trust will have a stronger positive relationship with perceived usefulness of an AI chatbot than system-like trust.

Perceived usefulness has consistently been identified as one of the strongest predictors of technology adoption in TAM and its extensions [107–109,116]. More recent work has emphasized that users who trust a system as reliable and capable are more likely to view it as valuable and useful, creating a positive relationship between trust and perceived usefulness. This is particularly when technologies are novel or involve uncertainty. Prior studies have shown that both human-like and system-like trust can positively influence usefulness judgments [24,117]. However, because chatbots engage students through conversational exchanges that resemble interpersonal communication, perceptions of ability and benevolence may carry more weight in usefulness evaluations. Based on this reasoning, we expect human-like trust to be more influential in explaining variance in perceived usefulness.

As mentioned above, while these hypotheses emphasize the expected salience of human-like trust across outcomes, it is also possible that its influence may vary depending on whether the outcome is emotional (e.g., enjoyment, trusting intention) or functional (e.g., usefulness, behavioral intention) [26,105]. Thus, finding stronger effects of human-like trust across all four outcomes would support its contextual salience for AI chatbots, whereas inconsistent patterns might signal limitations in existing frameworks. Such discrepancies could reflect measurement issues, but they may also point to deeper theoretical gaps and risk misrepresenting how trust operates in human-AI interactions.

## 4. Methods

### 4.1. Procedures and participants

An online survey was conducted with undergraduate students at the University of Florida through Qualtrics software during the Fall 2024 semester. Participants first reviewed and agreed to an informed consent document before beginning the study. After providing consent, participants reported demographic and background information, including age, gender, grade-point average (GPA), major, year in university, and previous experience with AI. To ensure participants shared a common frame of reference regarding the technology under study, all respondents viewed a short introductory video presenting the concept of an AI chatbot (i.e., ChatGPT) and an example interaction. This step was intended as a standardization tool to align participants with our definition and reference point for an AI chatbot, given the variability in students' prior exposure to such systems. Participants then responded to 28 randomized Likert-scale questions measuring trust (both human-like and system-like) and the four outcome variables (PE, TI, BI, PU).

The survey received 293 total responses, with 229 responses remaining for analysis after filtering for completeness. The final sample size exceeds the minimum requirements for PLS-SEM analysis, which should be equal to the larger of either: (1) ten times the largest number of formative indicators used to measure a construct or (2) ten times the largest number of structural paths directed at a particular construct in the structural model [118].

Participants reported high familiarity with AI chatbots, with 90.32 % indicating prior experience and a mean rating of 4.44 out of 5 on experience measures. In terms of frequency, 63.31 % were high-frequency users (daily or weekly use), 31.00 % were low-frequency users (monthly or less often), and only 5.68 % reported not using AI chatbots at all. The most common usage pattern was three to five times per week (27.07 %). The sample was 49.06 % female, 48.11 % male, and 2.83 % preferred not to disclose or self-describe their gender. The average age of participants was 20.27 years.

### 4.2. Measurement items

Measurement items were adapted from validated scales from prior research for all constructs. The second-order constructs, human-like trust (ability, benevolence, and integrity) and system-like trust (functionality, helpfulness, and reliability), were adapted from McKnight et al [[113], 2011]. The outcome variables included perceived enjoyment [119], trusting intention [113,120], behavioral intention [116,121], and perceived usefulness [107].

Because these instruments were originally developed in other technology and organizational contexts, items were adapted to reference AI chatbots in a learning environment. The present study therefore not only applies but also provides validation evidence for these adapted measures. All items were randomized within the survey and measured on a 5-point Likert scale ranging from *Strongly Disagree* (1) to *Strongly Agree* (5). The complete set of measurement items is provided in Appendix A.

### 4.3. Structural equation modeling

Partial least squares structural equation modeling (PLS-SEM) was conducted using SmartPLS 4 software [122]. PLS-SEM was chosen over covariance-based SEM (CB-SEM) due to its suitability for theory development and capability to deal with smaller sample sizes, complex models, and non-normally distributed data [118]. The analysis followed a two-stage approach following the guidance of Hair et al [118], first assessing the measurement model and then evaluating the structural model.

For the measurement model, we evaluated internal consistency reliability using composite reliability (CR), with values greater than 0.70 considered satisfactory [123]. We then assessed convergent validity using average variance extracted (AVE) with a threshold of 0.50. Discriminant validity was evaluated using both the Fornell-Larcker criterion, which compares the square root of AVE values with inter-construct correlations, and by checking cross-loadings between constructs [118,124]. For correlation interpretation, values between $\pm 0.50$ and $\pm 1.00$ indicate strong correlations, values between $\pm 0.30$ and $\pm 0.49$ represent moderate correlations, and values between $\pm 0.10$ and $\pm 0.29$ represent weak correlations, consistent with Cohen's [125] benchmarks.

For the structural model assessment, following Keith [126], we interpreted standardized path coefficients of 0.05, 0.10, and 0.25 as indicating small, moderate, and large effects respectively. Path coefficients' significance was tested through a bootstrapping procedure with 5000 samples, consistent with [127].

## 5. Results

### 5.1. Assessment of measurement model

The measurement model demonstrated strong reliability and validity across all constructs. Internal consistency reliability was confirmed through both Cronbach's alpha and composite reliability (CR) measures. Cronbach's alpha coefficients ranged from 0.823 to 0.947, exceeding the recommended threshold of 0.70. Similarly, CR values ranged from 0.883 to 0.962, well above the 0.70 threshold, further confirming measurement reliability.

Convergent validity was established through examination of factor loadings and average variance extracted (AVE). Factor loadings were robust across all items, with most exceeding 0.70, indicating strong item reliability. AVE values for all constructs surpassed the 0.50 threshold, ranging from 0.654 to 0.863, demonstrating strong convergent validity (see Table 1).

#### 5.1.1. Second-Order factor reliability & internal consistency

Reliability and internal consistency of the second-order constructs in the measurement model was confirmed, with all constructs achieving CR values greater than 0.70 and AVE values above 0.50 (See Table 2). The human-like trust construct demonstrated strong loadings with ability ($\beta = 0.887$, $p < 0.001$), benevolence ($\beta = 0.844$, $p < 0.001$), and integrity ($\beta = 0.829$, $p < 0.001$). Similarly, the system-like trust construct demonstrated strong loadings with functionality ($\beta = 0.933$, $p < 0.001$), helpfulness ($\beta = 0.949$, $p < 0.001$), and reliability ($\beta = 0.830$, $p < 0.001$).

#### 5.1.2. Discriminant validity and construct correlations

To assess discriminant validity, which indicates the extent to which constructs are distinct from each other, we first confirmed convergent validity, which serves as a prerequisite [128,129], and ensured that each item loaded uniquely on only one construct [129,130]. We then examined the Fornell-Larcker criterion (Table 3). Correlations were all strong (0.644 to 0.850), and the square root of AVE values for the outcome variables generally exceeded their correlations with other constructs in their respective rows and columns, indicating discriminant validity following the Fornell-Larcker criterion [124]. We observed higher correlations between system-like trust and other constructs (PU: 0.838, TI: 0.821) than the square root of system-like trust's AVE (0.782), which

**Table 1**
Evaluation of survey instrument: Item-level.

| Construct | Item | Factor Loading | M (SD) | Cronbach's α | CR (rho c) | AVE |
|---|---|---|---|---|---|---|
| Perceived Enjoyment (PE) | PE1 | 0.906 | 3.28 (1.14) | 0.916 | 0.940 | 0.798 |
| | PE2 | 0.859 | 3.18 (1.23) | | | |
| | PE3 | 0.923 | 3.25 (1.11) | | | |
| | PE4 | 0.884 | 3.58 (1.09) | | | |
| Trusting Intention (Trust) | Trust1 | 0.892 | 3.37 (1.16) | 0.922 | 0.945 | 0.811 |
| | Trust2 | 0.916 | 3.45 (1.13) | | | |
| | Trust3 | 0.912 | 3.48 (1.18) | | | |
| | Trust4 | 0.882 | 3.23 (1.16) | | | |
| Behavioral Intention (BI) | BI1 | 0.925 | 3.72 (1.13) | 0.947 | 0.962 | 0.863 |
| | BI2 | 0.918 | 3.73 (1.19) | | | |
| | BI3 | 0.932 | 3.91 (1.11) | | | |
| | BI4 | 0.941 | 3.85 (1.15) | | | |
| Perceived Usefulness (PU) | PU1 | 0.915 | 3.82 (1.08) | 0.906 | 0.935 | 0.781 |
| | PU2 | 0.828 | 3.46 (1.15) | | | |
| | PU3 | 0.886 | 3.63 (1.05) | | | |
| | PU4 | 0.905 | 3.82 (1.03) | | | |
| Ability (Abl) | Abl1 | 0.939 | 3.56 (1.03) | 0.943 | 0.959 | 0.855 |
| | Abl2 | 0.946 | 3.59 (1.09) | | | |
| | Abl3 | 0.948 | 3.66 (1.08) | | | |
| | Abl4 | 0.945 | 3.55 (1.10) | | | |
| Benevolence (Ben) | Ben1 | 0.891 | 3.22 (1.20) | 0.824 | 0.884 | 0.655 |
| | Ben2 | 0.806 | 3.73 (1.01) | | | |
| | Ben3 | 0.798 | 3.05 (1.20) | | | |
| | Ben4 | 0.732 | 3.26 (1.22) | | | |
| Integrity (Int) | Int1 | 0.882 | 3.04 (1.15) | 0.853 | 0.901 | 0.695 |
| | Int2 | 0.792 | 3.02 (1.13) | | | |
| | Int3 | 0.791 | 2.83 (1.21) | | | |
| | Int4 | 0.880 | 3.11 (1.12) | | | |
| Functionality (Func) | Func1 | 0.938 | 4.00 (0.97) | 0.898 | 0.929 | 0.767 |
| | Func2 | 0.933 | 3.77 (1.03) | | | |
| | Func3 | 0.900 | 3.90 (0.99) | | | |
| | Func4 | 0.862 | 3.57 (1.11) | | | |
| Helpfulness (Help) | Help1 | 0.944 | 3.90 (1.13) | 0.915 | 0.941 | 0.801 |
| | Help2 | 0.931 | 3.87 (1.12) | | | |
| | Help3 | 0.861 | 3.49 (1.18) | | | |
| | Help4 | 0.927 | 4.00 (1.15) | | | |
| Reliability (Rel) | Rel1 | 0.897 | 3.48 (1.12) | 0.823 | 0.883 | 0.654 |
| | Rel2 | 0.717 | 2.58 (1.19) | | | |
| | Rel3 | 0.893 | 3.25 (1.13) | | | |
| | Rel4 | 0.687 | 3.15 (1.15) | | | |

**Table 2**
Evaluation of survey instrument: Second-order constructs.

| Second-Order Construct | Construct | Loading | Cronbach's Alpha | CR (rho c) | AVE |
|---|---|---|---|---|---|
| Human-like Trust | Ability | 0.887 | 0.921 | 0.933 | 0.538 |
| | Benevolence | 0.844 | | | |
| | Integrity | 0.829 | | | |
| System-like Trust | Functionality | 0.933 | 0.940 | 0.949 | 0.612 |
| | Helpfulness | 0.949 | | | |
| | Reliability | 0.830 | | | |

**Table 3**
Discriminant validity: Square root of AVE (diagonal in bold) and correlations between constructs.

| | Hum-T | Sys-T | PE | TI | BI | PU |
|---|---|---|---|---|---|---|
| Hum-T | 0.733 | | | | | |
| Sys-T | 0.823 | 0.782 | | | | |
| PE | 0.686 | 0.692 | 0.893 | | | |
| TI | 0.850 | 0.821 | 0.712 | 0.900 | | |
| BI | 0.644 | 0.760 | 0.730 | 0.712 | 0.929 | |
| PU | 0.775 | 0.838 | 0.754 | 0.827 | 0.797 | 0.884 |

Note: Hum-*T* = Human-like Trust; Sys-*T* = System-like Trust; PE = Perceived Enjoyment; TI = Trusting Intention; BI = Behavioral Intention to use; PU = Perceived Usefulness.

suggests preliminary theoretical relationships between these constructs. However, we also noted that discriminant validity was not established between human-like trust and system-like trust ($r = 0.823$), as their correlation exceeded the square root of AVE, suggesting these measurement approaches may conceptually overlap when applied to student trust in AI systems. We address the implications of this finding in the discussion section.

### 5.2. *Structural model*

This study tested four hypotheses examining whether human-like trust would have stronger positive relationships than system-like trust with perceived enjoyment (H1), trusting intention (H2), behavioral intention to use (H3), and perceived usefulness (H4) of an AI chatbot.

#### 5.2.1. *Structural model assessment*

The structural model was assessed through analysis and comparison of the two second-order factors, human-like and system-like trust. PLS-SEM revealed significant relationships across 7 out of 8 paths, with results that both supported and contradicted our hypotheses (Table 5).

Human-like trust showed varied significance, with significant effects on perceived enjoyment ($\beta = 0.359$, $p < 0.001$), trusting intention ($\beta = 0.539$, $p < 0.001$), and perceived usefulness ($\beta = 0.266$, $p < 0.01$). One path from the human-like trust model failed to reach significance: behavioral intention ($\beta = 0.056$, $p = 0.520$). System-like trust had significant effects across all examined paths: perceived enjoyment ($\beta = 0.397$, $p < 0.001$), trusting intention ($\beta = 0.378$, $p < 0.001$), behavioral intention (β = 0.714, $p < 0.001$), and perceived usefulness (β = 0.618, $p < 0.001$).

Comparing the relative influence of human-like and system-like trusting beliefs, human-like trust demonstrated similar effects on perceived enjoyment (H1; β = 0.359 vs. β = 0.397) and stronger effects on trusting intention (H2; β = 0.539 vs. β = 0.378), while system-like trust showed stronger effects on behavioral intention (H3; β = 0.714 vs. β = 0.056), and perceived usefulness (H4; β = 0.618 vs. β = 0.266). The results of the PLS-SEM analysis are presented in Table 4, with each hypothesis outcome summarized in Table 5, and visualized in Fig. 2.

## 6. Discussion

With the growing adoption of AI chatbots in education, it is important to understand how trust in these systems should be measured and modeled. The extent to which students place trust in AI-enabled educational technologies has influence on whether students engage in productive behaviors such as critical thinking and sustained interaction, or fall into disengagement and over-reliance that undermine learning.

We sought to understand if students develop trust in AI chatbots similar to how they trust human peers or instructors (through human-like trusting beliefs), or as they would develop trust in typical software applications (through system-like trusting beliefs). Our findings reveal that both human-like and system-like trusting beliefs significantly influence student perceptions, though their effects vary by outcome. The mixed pattern of outcomes suggests that human-AI trust cannot be fully explained by either interpersonal or technology trust models alone. Instead, it appears to draw from both, creating conceptual overlap that could indicate either a distinct hybrid form of trust or a two-factor solution in which human-like and system-like trust operate together but remain partially separable. Clarifying whether human-AI trust should be modeled as a single construct or as two interacting dimensions is therefore an open question. Based on these findings, the following sections discuss the theoretical implications for how trust in AI might be conceptualized, as well as practical implications for designing and implementing AI systems in ways that foster calibrated trust that supports rather than hinders learning.

### 6.1. *Theoretical implications*

Our findings demonstrate asymmetric effects of existing trust frameworks on students' perceptions of and behavioral intentions toward AI chatbots, carrying important theoretical implications. Human-like trust showed stronger influence on trusting intention and similar effects on enjoyment, while system-like trust was more predictive of behavioral intention and perceived usefulness. This challenged our initial hypothesis that human-like trust would be universally dominant, suggesting instead that each framework relates to different aspects of students' perceptions and trust in AI.

These findings indicate limitations in current measurement approaches. When trust is measured only through human-like or only through system-like beliefs, as is traditionally done, they may reach different conclusions about trust's influence on outcomes like behavioral intention or perceived usefulness - not because trust itself differs, but because each approach captures a partial view. Consequently, findings risk becoming artifacts of measurement choice rather than accurate reflections of student-AI interactions.

**Table 4**
Results of the PLS-SEM Analysis.

| Path | Path Coef. | t-value | p-value |
|---|---|---|---|
| Hum-T → PE | 0.359 | 3.990 | 0.000 |
| Hum-T → Trust | 0.539 | 8.710 | 0.000 |
| Hum-T → BI | 0.056 | 0.643 | 0.520 |
| Hum-T → PU | 0.266 | 3.144 | 0.002 |
| Sys-T → PE | 0.397 | 4.579 | 0.000 |
| Sys-T → Trust | 0.378 | 6.122 | 0.000 |
| Sys-T → BI | 0.714 | 8.942 | 0.000 |
| Sys-T → PU | 0.618 | 7.837 | 0.000 |

Note: Hum-*T* = Human-like Trust; Sys-*T* = System-like Trust; PE = Perceived Enjoyment; TI = Trusting Intention; BI = Behavioral Intention to use; PU = Perceived Usefulness.

**Table 5**
Hypothesis testing.

| Hypothesis | Outcome Variable | Predicted Direction | Supported? |
|---|---|---|---|
| H1 | Perceived Enjoyment | Human-like > System-like | No |
| H2 | Trusting Intention | Human-like > System-like | Yes |
| H3 | Behavioral Intention | Human-like > System-like | No |
| H4 | Perceived Usefulness | Human-like > System-like | No |

Our discriminant validity analysis provides additional evidence for such measurement limitations identified above. The Fornell-Larcker criterion revealed that system-like trust's square root of AVE (0.782) fell below its correlation with human-like trust (0.823), indicating conceptual overlap between these models when applied to student-AI trust, a pattern not observed in prior applications of the human-like/system-like dichotomy for database software [24], Facebook [24], Airbnb [25], and mobile banking applications [106]. Whether the anthropomorphic characteristics of AI chatbots drive this overlap as hypothesized, or additional situational factors, the lack of discriminant validity observed indicates that AI technologies require more contextual trust models that account for the unique relationships individuals develop with these systems. Future work should focus on developing refined models of human-AI trust that capture the affective, relational, and functional dimensions unique to these interactions, extending beyond existing human-human and human-technology trust theory.

At the same time, our structural results suggest a more nuanced interpretation. Even though the second-order constructs are empirically entangled, they still predict outcomes differently, with system-like trust driving usefulness and behavioral intention, and human-like trust driving trusting intention. This divergence suggests that the constructs are not redundant but may operate as interacting pathways within a broader trust process. From this perspective, the overlap does not invalidate the two-factor model but points to the need for future work to extend it for contexts like AI chatbots, where technologies present both functional and social affordances and where models must account for their joint influence on trust. In this sense, future work may also reconceptualize human-like and system-like trust not as mutually exclusive dimensions but as jointly activated mechanisms that converge into a hybrid form of trust specific to human-AI interaction.

Toward an understanding of human-AI trust in student-AI interactions, the varying effects observed in the structural model also suggest that AI chatbots' functional capabilities (e.g., to output explanations, feedback, or learning resources) drive students' perceptions of usability (i.e., usage intentions and perceived usefulness), while social affordances foster human-like trust that more strongly influences emotionally-based responses (i.e., willingness to depend on the system and perceived enjoyment). This extends the research questions posed by Lankton et al [24] by demonstrating that the anthropomorphic characteristics of AI chatbots may determine not just which trust framework is more influential overall, but how both human-like and system-like trust can be simultaneously significant, having greater influence on different outcomes. Additionally, this result aligns with Bhatti and Robert's [131] observation that anthropomorphism is often a vague and cognitively straining process, as people grapple with whether human-like qualities stem from design cues or from their own attributions during interaction. This ambiguity can produce category confusion and strain in interpreting intelligent systems. Similarly, Endsley [132] identifies five "ironies of AI," including the difficulty of understanding and adapting to AI systems, the opaqueness of their limitations and biases that can influence human decision biases, and challenges in assessing the reliability of AI tools. These perspectives suggest that the very features making AI appear more human-like or sophisticated can at the same time complicate users' ability to form accurate and calibrated trust.

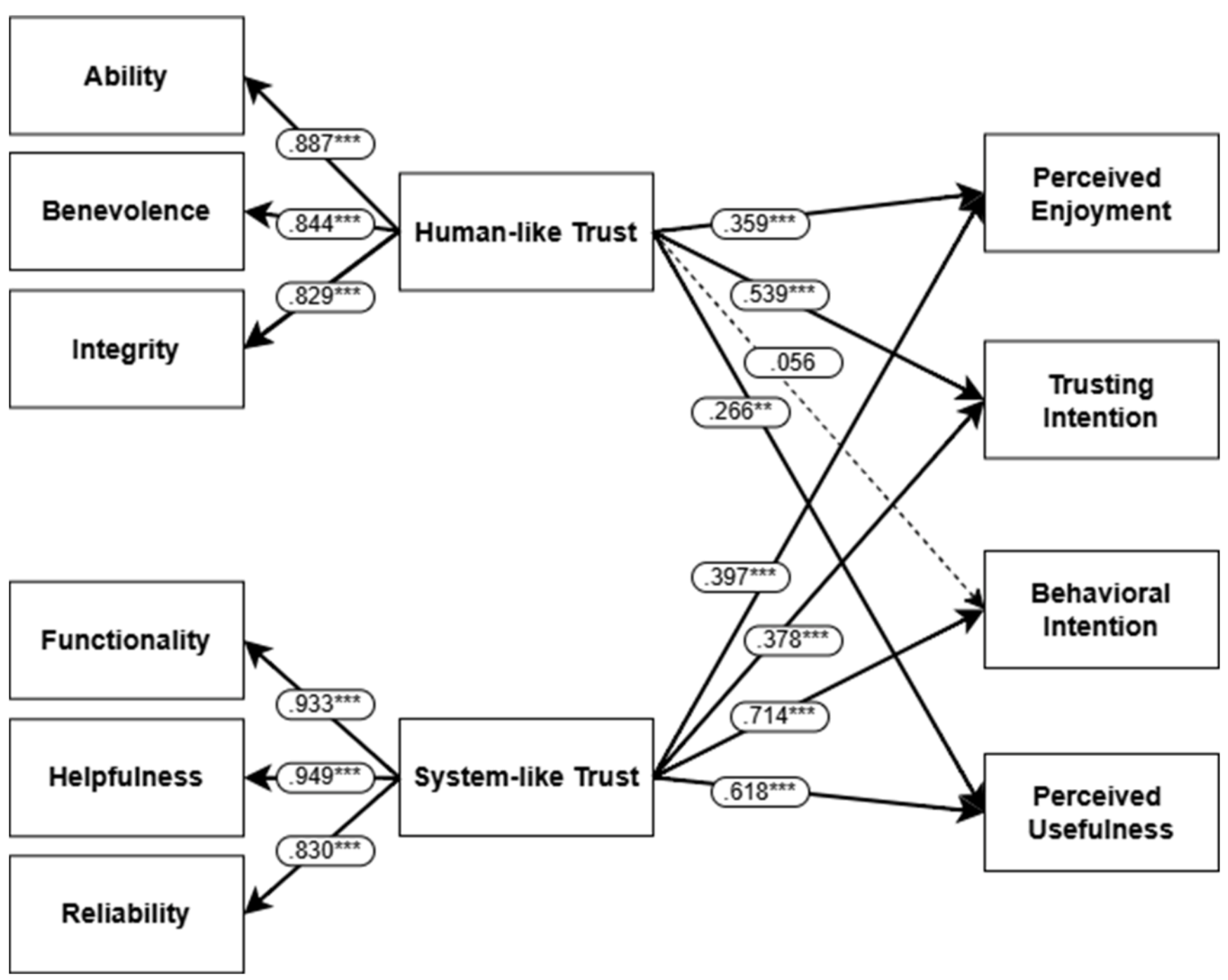

**Fig. 2.** Assessment of the structural model. Dashed lines indicate a non-significant path. Standardized path coefficients. $*p < 0.05$, $**p < 0.01$, $***p < 0.001$.

We also note that contextual factors, in our case the educational context, may amplify this effect. Students may prioritize functionality and reliability when deciding whether to use and trust AI tools for learning, while at the same time engaging with them conversationally in ways that resemble interactions with instructors or peers. Future research should explore how different contextual factors influence trusting relationships and investigate the relationship between human-like and system-like trust across a broader range of AI applications toward a better understanding of human-AI trust.

To avoid reducing human-AI trust as a variant of either human-human or human-technology trust, we define it as *a distinct construct that reflects confidence in an AI system's capacity to act intelligently, reliably, and in alignment with the user's goals under uncertainty*. Unlike interpersonal trust, which presumes intentionality, or technology trust, which emphasizes stable functionality, human-AI trust is shaped by how people interpret an artificial system's decision-making processes (e.g., accuracy, explainability, adaptability) alongside its social presentation (e.g., conversational dialogue, responsiveness). We propose that human-AI trust is defined by three features: (1) reliance on AI's computational intelligence rather than human intentionality, (2) evaluation of the quality of outputs and the quality of interactions, and (3) sensitivity to uncertainty, as AI systems generate outputs probabilistically and may produce errors even when appearing confident. Framed in this way, human-AI trust captures the distinctive ways people engage with and trust intelligent systems in education and across other domains where AI is applied. This framing also lays the groundwork for practical implications, as users' formation of trust in AI systems directly influences how they calibrate and apply that trust in practice. Next, we focus on the practical implications in the context of education.

### 6.2. Practical implications

The results of this study offer several practical implications for developers and educators First, to maximize perceptions of AI systems' usefulness and subsequent engagement, developers should prioritize system reliability, functionality, and helpfulness, as system-like trust shows a stronger relationship with perceived usefulness ($\beta = 0.618$, $p < 0.001$) than human-like trust ($\beta = 0.266$, $p < 0.01$). The low mean ratings on reliability items, Rel2 ($M = 2.58$) and Rel3 ($M = 3.25$), compared to functionality items like Func1 ($M = 4.00$), reveal that students currently see a gap between AI tools' potential and their actual dependability. This reliability gap may lead to inconsistent use patterns that undermine the effectiveness of AI-enabled learning systems. Given this, developers should focus on improving system consistency, providing clearer information about when systems might fail, and implementing better error handling to address these reliability concerns.

Second, to build students' willingness to depend on AI systems for learning support, developers should address concerns about system integrity and benevolence alongside functional improvements. Human-like trust strongly influences trusting intention ($\beta = 0.539$, $p < 0.001$), and the low ratings for integrity items like Int3 ($M = 2.83$) and benevolence items like Ben3 ($M = 3.05$) suggest areas for improvement. When students doubt whether AI systems prioritize their learning needs or provide truthful information, they may engage superficially with the technology, limiting potential learning benefits. Developers can foster students' trust by implementing transparent information sourcing, clearly communicating how AI responses are generated, acknowledging system limitations upfront, and demonstrating through design choices that the system prioritizes student learning over other objectives.

Third, the findings provide insights for addressing concerns relating to potentially excessive trust and distrust, to help students develop appropriately calibrated trust. The gap between students' recognition of AI utility (PU1, $M = 3.82$) and their concerns about system integrity (Int3, $M = 2.83$) suggests they understand potential benefits while maintaining healthy skepticism about limitations. This imbalance with higher trust in functionality/ability and higher distrust in integrity/reliability presents an opportunity for targeted trust interventions to calibrate students' trust and optimize their engagement. Rather than trying to maximize trust universally, developers should design systems that support both effective use and critical evaluation of outputs. Such supports could include explainable outputs, confidence indicators, or mechanisms that highlight when responses may be uncertain, complemented by educational resources that help students interpret these signals and recognize appropriate use cases. Building on earlier interventions aimed at reliability, benevolence, and integrity, these additional features can better equip students to calibrate trust and engage with AI in ways that enhance, rather than compromise, their learning outcomes.

Finally, these findings have direct pedagogical implications. Because students' behavioral intentions are driven more by system-like trust while their willingness to depend is driven more by human-like trust, instructors and institutions must carefully consider how they introduce and frame AI chatbots in educational contexts. The challenge is to strike a balance between encouraging students to make productive use of these systems and ensuring they maintain a critical stance toward the outputs they generate. If students place too much weight on system-like trust, they may come to view chatbot responses as authoritative, adopting them uncritically and substituting AI answers for their own reasoning. This kind of over-reliance risks not only weakening individual problem-solving skills but also raising questions of authorship and academic integrity. Conversely, if students rely only on human-like trust, they may treat chatbots as conversational partners or sources of affirmation, engaging socially without sufficiently interrogating the accuracy or appropriateness of the information provided. Both risk undermining the educational value of these tools.

To help students balance these tendencies, instructors can position chatbots as supplementary tools, rather than as replacements for independent thought. This framing can be reinforced through classroom discourse as well as through intentional course design. For example, instructors might create assignments where students must use chatbot outputs as a starting point and then refine, critique, or expand upon them. Verification exercises, in which students compare chatbot responses against authoritative sources such as textbooks or peer-reviewed material, can train students to approach outputs skeptically and to value evidence-based reasoning. Reflection activities can also be incorporated, prompting students to consider when reliance on AI is appropriate, when it should be limited, and how their trust in the system may shift across contexts.

Institutions can further support the calibration of appropriate levels of trust by embedding AI literacy into curricula. This curriculum should extend beyond technical familiarity with how to operate AI systems. Instead, it should include explicit instruction about the ethical use of AI, such as understanding potential biases in outputs, recognizing the limits of generative models, and grappling with questions of fairness, transparency, and accountability. Students should also be provided with clear guidance on appropriate reliance in different learning contexts, for example, distinguishing between using AI as an exploratory brainstorming tool versus relying on it for solutions in high-stakes settings. In addition, institutions play an important role in reinforcing academic integrity norms and clarifying policies around acceptable AI use. Importantly, adopting blanket prohibitions on AI can have unintended consequences, such as limiting opportunities for students to learn how to engage with these tools responsibly and leaving them unprepared for settings where AI is widely used. Clear and balanced policies that encourage responsible use are more likely to promote academic integrity and help students develop the judgment to determine when and how AI should be relied upon.

Together, these practical and pedagogical implications emphasize that the challenge is not to maximize trust in AI but to calibrate it appropriately. By strengthening reliability and transparency, supporting both effective use and critical evaluation of outputs, and embedding AI literacy into curricula, developers and educators can help students engage productively with AI systems. When trust is calibrated in this way, AI can enhance critical thinking, sustain academic integrity, and contribute to learning outcomes while supporting, rather than substituting for, student engagement.

## 7. Limitations

This study has limitations that can be addressed through further research. First, the sample consisted of undergraduate students from a single university surveyed in Fall 2024. This may constrain the applicability of our model and findings, as students' experiences in one institutional context may not generalize to learners at other institutions, to non-student populations, or to educational settings with different pedagogical approaches. Second, participants' prior experiences with AI varied, and although we provided an introductory video to create a shared point of reference, students' prior interactions may have had varying influence on their perception of the AI's capabilities and limitations. Moreover, the relationship between students' reported and actual trust in extended interactions remains unclear; this is a limitation of much current trust research, and the precise measurement of trust in relationships is a current research interest and gap in the field. Further, our study did not account for individual differences such as prior AI literacy, technical self-efficacy, or personal learning preferences, which may moderate the relationships we observed. Lastly, our focus on chatbot-based AI learning assistants represents just one implementation of AI in education. The relative influence of human-like and system-like trust may vary across different AI technologies with varying degrees of embodiment or pedagogical roles.

## 8. Conclusion

This study was conducted to understand the relative importance of human-like and system-like trusting beliefs in students' perceptions of AI chatbots in education, with implications for maximizing the pedagogical potential of these tools. Using partial least squares structural equation modeling on survey data from 229 undergraduate students, we identified asymmetric patterns of influence: human-like trust more strongly predicted trusting intention and enjoyment, while system-like trust was more influential for behavioral intention and perceived usefulness. These results challenged our initial hypothesis that human-like trust would be more influential across outcomes. They also point toward the need for theoretical frameworks that account for how functional, social, and uncertain qualities of AI systems jointly influence individuals' formation of trust.

The findings carry theoretical, practical, and pedagogical implications. Theoretically, they reveal limitations of established interpersonal and technology trust frameworks, with conceptual overlap indicating the need for new models tailored to human-AI interaction. Practically, gaps in students' perceptions of AI benevolence, reliability, and integrity highlight opportunities for developers to enhance transparency, strengthen reliability, and design features that foster critical evaluation of outputs. Pedagogically, instructors and institutions can support calibrated trust by embedding AI literacy into curricula, positioning chatbots as supplementary aids rather than substitutes, and clarifying policies around appropriate reliance and academic integrity.

As AI systems become more prevalent in education, advancing theoretical models of human-AI trust and translating them into design and instructional practices will be critical for ensuring that these technologies enhance learning outcomes and student experiences responsibly.

## Statements on ethics

This study was approved by the Institutional Review Board at the University of Florida (Protocol #ET00024199). All participants provided informed consent before participating in the study.

## CRediT authorship contribution statement

**Griffin Pitts:** Writing – review & editing, Writing – original draft, Visualization, Software, Methodology, Investigation, Formal analysis,

Data curation, Conceptualization. **Sanaz Motamedi:** Writing – review & editing, Writing – original draft, Supervision, Project administration, Methodology.

### Declaration of competing interest

The authors declare that they have no known competing financial interests or personal relationships that could have appeared to influence the work reported in this paper.

### Acknowledgements

The authors acknowledge those who assisted in the collection of data for this study, including the survey participants, and instructors at the University of Florida, for sharing and completing the survey.

## Appendix A. Instrument Measures

| Construct | Item | Item Text | Reference |
| --- | --- | --- | --- |
| Perceived Enjoyment | PE1 | An AI chatbot would make learning more interesting. | Adapted from Davis et al [119] |
| | PE2 | An AI chatbot would make learning more engaging. | |
| | PE3 | An AI chatbot would make learning more enjoyable. | |
| | PE4 | I would enjoy using an AI chatbot as a learning aid. | |
| Trusting Intention | Trust1 | I could depend on an AI chatbot for assistance while I am learning. | Adapted from McKnight et al., [113]; Gulati et al., [120] |
| | Trust2 | I could rely on an AI chatbot for assistance while I am learning. | |
| | Trust3 | In general, I could count on an AI chatbot for assistance while learning. | |
| | Trust4 | I could trust an AI chatbot for assistance while I am learning. | |
| Behavior Intention | BI1 | I would use an AI chatbot while I am learning. | Adapted from Venkatesh et al., [116]; Davis et al [121] |
| | BI2 | I plan to use an AI chatbot while learning in the future. | |
| | BI3 | If given permission to use an AI chatbot for an academic course, I would use it. | |
| | BI4 | If given permission to use an AI chatbot for an academic course, I predict that I would use it. | |
| Perceived Usefulness | PU1 | An AI chatbot would be a useful tool to have while learning. | Adapted from Davis et al [121] |
| | PU2 | An AI chatbot would decrease my frustration while learning. | |
| | PU3 | An AI chatbot would provide valuable academic support when I require it. | |
| | PU4 | An AI chatbot would be of use to me while I am learning. | |
| Ability | Abl1 | An AI chatbot would be a competent and effective assistant as a learning aid. | Adapted from McKnight et al., [113] |
| | Abl2 | An AI chatbot would be a capable and proficient tool as a learning aid. | |
| | Abl3 | An AI chatbot would assist as a learning aid well. | |
| | Abl4 | An AI chatbot would be a skillful assistant as a learning aid. | |
| Benevolence | Ben1 | An AI chatbot would act in my best interest as a learning assistant. | Adapted from McKnight et al., [113] |
| | Ben2 | An AI chatbot would do its best to help me if I needed assistance while learning. | |
| | Ben3 | An AI chatbot would prioritize my learning needs over itself as a learning assistant. | |
| | Ben4 | An AI chatbot would do its best to be kind and thoughtful while assisting me. | |
| Integrity | Int1 | An AI chatbot would be truthful in a conversation with me. | Adapted from McKnight et al., [113] |
| | Int2 | An AI chatbot would keep its commitments. | |
| | Int3 | An AI chatbot would be sincere while interacting with others. | |
| | Int4 | An AI chatbot would be honest in conversations. | |
| Functionality | Func1 | AI chatbots have the functionality to provide me assistance while I am learning. | Adapted from McKnight et al [22] |
| | Func2 | AI chatbots have the required features to help me while I am learning. | |
| | Func3 | AI chatbots have the capability to provide the assistance I require of them while I am learning. | |
| | Func4 | AI chatbots have the functionality to do what I want them to do. | |
| Helpfulness | Help1 | An AI chatbot would help me as I am learning. | Adapted from McKnight et al [22] |
| | Help2 | An AI chatbot would provide me with helpful guidance as I am learning. | |
| | Help3 | An AI chatbot would provide me with whatever assistance I need as I am learning. | |
| | Help4 | In general, an AI chatbot would be helpful as a learning aid. | |
| Reliability | Rel1 | An AI chatbot would be a reliable learning aid. | Adapted from McKnight et al [22] |
| | Rel2 | An AI chatbot would not fail to assist me. | |
| | Rel3 | An AI chatbot would be a dependable learning assistant. | |
| | Rel4 | An AI chatbot would be consistent in a conversation with me. | |

### Data availability

The anonymized data that support the findings of this study are available from the corresponding author, GP, upon reasonable request.